\documentclass[aps,amsmath,amssymb,superscriptaddress,twocolumn,color,epsfig,graphicx,bm,floatfix]{revtex4-1}

\usepackage{xcolor}
\usepackage{makecell}
\usepackage{graphicx}
\usepackage[caption=false]{subfig}
\usepackage{epstopdf}
\usepackage{tabularx}
\usepackage[super]{nth}
\usepackage[caption=false]{subfig}
\usepackage{amsmath}    
\usepackage{mathtools}
\usepackage{siunitx}
\usepackage{multirow}
\usepackage{array}
\usepackage[version=4]{mhchem}
\newcolumntype{K}[1]{>{\centering\arraybackslash}p{#1}}

\usepackage{verbatim}   
\usepackage{color}      
\usepackage{hyperref}   
\begin{document}


\author{Duncan R. Sutherland}
\thanks{These authors contributed equally to this work}
\affiliation{Department of Materials Science and Engineering, Cornell University, Ithaca, NY 14853, United States}

\author{Aine Boyer Connolly}
\thanks{These authors contributed equally to this work}
\affiliation{Department of Materials Science and Engineering, Cornell University, Ithaca, NY 14853, United States}

\author{Maximilian Amsler}
\affiliation{Department of Materials Science and Engineering, Cornell University, Ithaca, NY 14853, United States}
\affiliation{Department of Chemistry and Biochemistry, University of Bern, Freiestrasse 3, CH-3012 Bern, Switzerland}

\author{Ming-Chiang Chang}
\affiliation{Department of Materials Science and Engineering, Cornell University, Ithaca, NY 14853, United States}

\author{Katie Rose Gann}
\affiliation{Department of Materials Science and Engineering, Cornell University, Ithaca, NY 14853, United States}

\author{Vidit Gupta}
\affiliation{Department of Materials Science and Engineering, Cornell University, Ithaca, NY 14853, United States}

\author{Sebastian Ament}
\affiliation{Department of Computer Science, Cornell University, Ithaca, NY 14853, United States}

\author{Dan Guevarra}
\affiliation{Joint Center for Artificial Photosynthesis, California Institute of Technology, Pasadena, CA 91125}

\author{John M. Gregoire}
\affiliation{Joint Center for Artificial Photosynthesis, California Institute of Technology, Pasadena, CA 91125}

\author{Carla P. Gomes}
\affiliation{Department of Computer Science, Cornell University, Ithaca, NY 14853, United States}

\author{R. B. van Dover}
\email{rbv2@cornell.edu}
\affiliation{Department of Materials Science and Engineering, Cornell University, Ithaca, NY 14853, United States}

\author{Michael O. Thompson}
\email{mot1@cornell.edu}
\affiliation{Department of Materials Science and Engineering, Cornell University, Ithaca, NY 14853, United States}

\title[Methodology]
  {Optical Identification of Materials Transformations in Oxide Thin Films}



\date{\today}
\begin{abstract}
Recent advances in high-throughput experimentation for combinatorial studies have accelerated the discovery and analysis of materials across a wide range of compositions and synthesis conditions.  However, many of the more powerful characterization methods are limited by speed, cost, availability, and/or resolution.  To make efficient use of these methods, there is value in developing approaches for identifying critical compositions and conditions to be used as a-priori knowledge for follow-up characterization with high-precision techniques, such as micron-scale synchrotron based X-ray diffraction (XRD). Here we demonstrate the use of optical microscopy and reflectance spectroscopy to identify likely phase-change boundaries in thin film libraries. These methods are used to delineate possible metastable phase boundaries following lateral-gradient Laser Spike Annealing (lg-LSA) of oxide materials. The set of boundaries are then compared with definitive determinations of structural transformations obtained using high-resolution XRD. We demonstrate that the optical methods detect more than 95\% of the structural transformations in a composition-gradient La-Mn-O library and a \ce{Ga2O3} sample, both subject to an extensive set of lg-LSA anneals. Our results provide quantitative support for the value of optically-detected transformations as \emph{a priori} data to guide subsequent structural characterization, ultimately accelerating and enhancing the efficient  implementation of $\mu$m-resolution XRD experiments.

\end{abstract}
\maketitle

\section{Introduction}

Combinatorial materials science research employs high throughput experimentation to rapidly probe and analyze large regions of synthesis and processing space~\cite{Green2017,Vasudevan2019}. The exploration of metastable phases is of particular interest in the context of designing new materials~\cite{CNGMD} and synthesis of metastable materials requires precise control of transformation kinetics. Structural characterization during metal oxide synthesis has revealed a complex interplay between thermodynamics and kinetics as a function of composition, temperature, and time\cite{bianchini_interplay_2020}, which is consistent with the general understanding of phase behavior/evolution in materials. The composition-temperature-time synthesis phase space is rich in structure and largely unexplored due to its complexity and the limited ability to process materials on fast time scales.  Conventional methods for annealing thin film samples include hot plate, furnace, and rapid thermal annealing~\cite{Borisenko1997}, although the 1 - 100~K/s cooling rates available with these techniques are insufficient to kinetically trap many high-temperature metastable phases for room temperature characterization. Methods for studying metastable materials include characterization after quenching during thin film deposition, both with and without subsequent annealing\cite{ren_accelerated_2018,ruiz-yi_different_2016}, and nanocalorimetric processing that provides quench rates on the order of $10^4$~K/s\cite{gregoire_combining_2012}.

Recently, lateral gradient Laser Spike Annealing (lg-LSA) has been developed as a high-throughput technique in which thin film samples are annealed in the sub-millisecond~\cite{Bell2016} time domain. The specific Gaussian-like laser beam profile in lg-LSA generates a reproducible spatial temperature gradient across the scan direction, typically  $\sim1$ mm in width (FWHM). With an accessible domain of peak temperatures ($300-1410\ ^\circ$C) and dwell/quench rates ($200~\mu\text{s}-10~\text{ms}$), a wide range of anneal conditions can be probed rapidly in a single thin film sample. We have demonstrated that lg-LSA can form and quench metastable phases in functional oxide materials, for example, by stabilizing high-temperature polymorphs at ambient conditions~\cite{Bell2016}. Due to the large thermal gradient across an lg-LSA stripe, $\mu$m-scale spatial resolution is required to analyze the processed thin film properties. Structural characterization using X-ray diffraction (XRD) allows the identification of the cyrstalline phases present in addition to providing information on the overall degree of crystallinity\cite{He2009}. However, typical lab-source diffractometers are unsuitable for resolving structural transformations across an lg-LSA stripe in a high-throughput fashion since they (a) produce spot sizes on the order of 1~mm or larger, and (b) provide insufficient X-ray intensity to account for the weak scattering of thin films, demanding long integration times. Using slits to improve spatial resolution of the X-ray source further degrades the throughput. On the other hand synchrotron facilities produce X-rays with very high intensity, appropriate wavelength, and $\mu$m-scale spatial extent, but are only available during limited beam times. The typically short duration of synchrotron access compared to a complete synthesis and processing cycle limits the ability to adapt the synthesis and processing strategy based on evolving knowledge of structural transitions.

To address these limitations, we have developed two complementary, spatially resolved optical analysis methods that permit rapid identification of the transitions occurring in lg-LSA annealed samples, data that then can guide follow-up, structural characterization: (a) white-light or narrow wavelength (e.g., LED) optical microscopy and (b) spectroscopic reflectometry. By comparing these methods to spatially-resolved XRD reference data for two materials systems (single-composition \ce{Ga2O3} and composition-gradient La-Mn-O) we demonstrate that the presence and location of phase transitions can be reliably identified using optical techniques prior to X-ray characterization. These readily accessible benchtop methods drastically improve the throughput of materials analysis, allowing precious synchrotron beam time to be used more effectively. Because they are readily available, our methods can speed up the iterative feedback cycle between characterization and synthesis/processing by orders of magnitude.

In Sec.~\ref{sec:methods} we discuss the methods. The two spatially resolved optical analysis methods developed in this work are described in detail in Sec.~\ref{sec:spatially}. Application of these methods to two materials systems is discussed in Sec~\ref{sec:results}, which are then assessed for accuracy using two metrics in  Secs~\ref{sec:RMSG} and~\ref{sec:completeness}. Finally, we summarize our results and conclude in Sec. ~\ref{sec:conclusions}.

\section{Methods\label{sec:methods}}
\subsection{Thin film deposition and lg-LSA processing}
Thin film libraries were deposited onto thermally oxidixed (20~nm) highly doped (\emph{p}-type, 0.02~$\Omega$ cm), lithographically patterned Si wafers with gold alignment marks. The \ce{Ga2O3} sample was deposited via reactive sputtering from a \ce{Ga2O3} target in an atmosphere of 4.5~mTorr Ar and 0.5mTorr \ce{O2} in an AJA sputter system. The target was operated at an RF power of 210 W while the substrate was rotated to create a 160~nm thick film with a $<5\%$ thickness variation.

The La-Mn-O system was deposited by reactive co-sputtering from La and Mn targets offset to generate a lateral composition gradient using a custom deposition system described previously~\cite{Suram2015}. In an atmosphere of 5.4 mTorr Ar and 0.6 mTorr \ce{O2}, the La and Mn targets were energized with an RF power of 83 W and 140 W, respectively.  XRF measurements, calibrated using thin film standards, revealed that La/(La+Mn) concentration in the as-deposited film varied from 12 to 78$\%$ across the library with a metals loading at substrate center of 1.4 micromol/cm$^2$, corresponding to a thickness of approximately 250 nm.

The beam of a \ce{CO2} laser operating at $\lambda = 10.6\mu$m and maximum power of 125~W was shaped to provide a $320~\mu$m-wide (full width at half maximum) lateral beam profile and used to generate lg-LSA stripes on the two materials system samples. Each anneal stripe was 5~mm long, with peak temperatures ranging from 600$^{\circ}$C to 1410$^{\circ}$C and processing dwell times between 250~$\mu$s and 10~ms. The dwell is defined as the FWHM of the laser in the scan direction divided by the scan velocity; this is approximately the time the temperature is within 5~\% of the peak temperature~\cite{Bell2016}. The anneal locations were randomized across the thin film library to avoid location bias on the wafer and to evenly distribute the time and temperature conditions with respect to the composition gradient in the La-Mn-O sample. A 100~mm diameter wafer offers space for a total of up to 625 stripes with distinct anneal conditions.  

\subsection{Spatially resolved techniques\label{sec:spatially}}

\subsubsection{Optical microscopy\label{sec:microscopy}}
Optical micrographs were collected using a Thorlabs CMOS camera (RGB channels with $1024 \times 1280$ pixels) aligned normal to the sample. The sample was illuminated coaxially with white light over a spot size approximately 1~mm in diameter. The R,G,B color channels were summed using the  weights of (0.3, 0.59, 0.11) and (0.33, 0.33, 0.33) for the \ce{Ga2O3} and La-Mn-O samples, respectively, to create high-contrast grayscale images. The magnification was set to provide a 0.79~$\mu$m spacing between pixels in the final image; this produced images with an approximately 1~mm horizontal field of view (FOV). A representative RGB micrograph is shown in Figure~\ref{fig:mapcreation}~a), which is horizontally cropped to be consistent with the displays associated with spectroscopic and diffraction techniques. 

\subsubsection{Reflectance spectroscopy\label{sec:spectroscopy}}

For reflectance spectroscopy, the sample was illuminated with a white light source ($400 <\lambda < 90$~nm). A seven linear fiber bundle was imaged onto the sample to collect reflected light from an area approximately $10\mu\text{m}\times 70\mu\text{m}$, aligned along the laser scan direction. The fiber bundle (a) improves the signal to noise ratio and (b) reduces erroneous signal from defects and particles. A reference dark spectrum and a spectrum from an Ag-coated mirror were used to calibrate the sample spectra. The fiber image was scanned across laser stripes at 10~$\mu$m increment for 201 samples, each with an optimized integration time of $\approx 250$~ms. Reflectance traces from three representative scans are shown in Figure \ref{fig:mapcreation}~c).

\subsubsection{X-ray diffraction}
The data from the two optical techniques were compared to reference data collected using the ID3B beamline at the Cornell High Energy Synchrotron Source (CHESS). A 9.7~keV beam was focused to yield a spot size with a $20\mu\text{m}\times40\mu\text{m}$ footprint on the sample at a 2$^\circ$ angle of incidence. The diffracted signal was captured with a Pilatus 300K detector. The X-ray beam was scanned along the same path as in the reflectance spectroscopy technique, collecting data every 10~$\mu$m across the anneal stripe with a 50~ms integration time. The 2D detector data was integrated along the $\chi$ direction using pyFAI~\cite{Ashiotis2015}. 

\subsection{Analysis}
Systematic data processing protocols were developed in order to compare the different techniques. Since they vary in their spatial extent and resolution, a key step involved carefully aligning the resulting data maps across the techniques to allow a quantitative comparison. The optical micrographs were cropped to conform to the dimensions of the region probed by the reflectance spectroscopy. Each map was smoothed with the same spatial degree of filtering to decrease noise. An unsharp mask filter was applied to the x-ray diffraction maps to enhance sharpness and emphasize featuring.

\section{Results and Discussion\label{sec:results}}

\begin{figure*}[t]
    \centering
    \includegraphics[width=\textwidth]{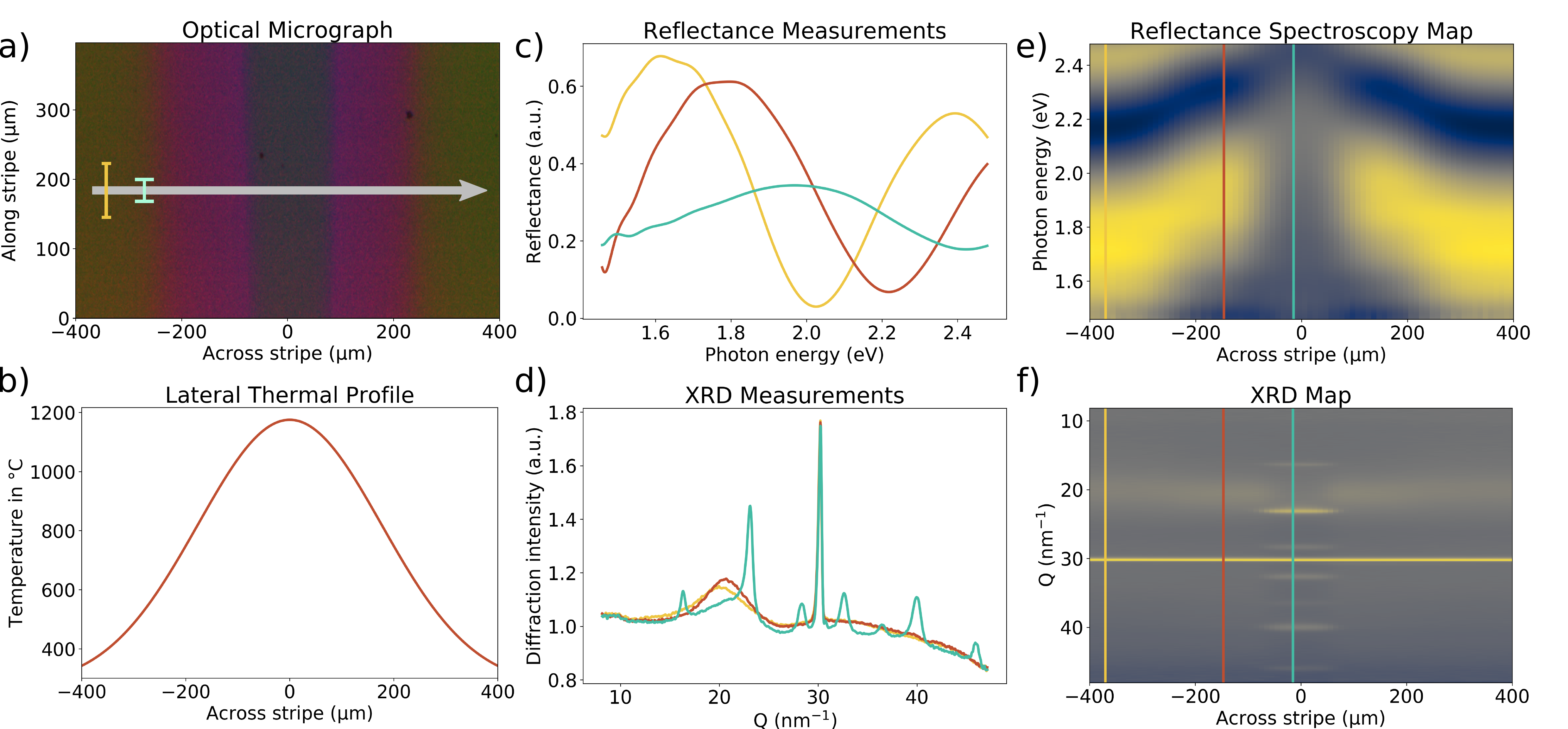}
    \caption{Acquisition and construction of the spatially resolved technique maps for a representative lg-LSA stripe (vertical scan) on the La-Mn-O sample, annealed at a peak temperature of 1175 $^{\circ}$C for 3000$\mu$s with a nominal La atomic fraction of 0.7. Due to the lateral profile of the incident laser, the center of the stripe corresponds to the peak temperature, decreasing smoothly to ambient on either side, panel b). Panel a) shows an optical micrograph with reference probe sizes for reflectance spectroscopy (yellow bar, $10\mu$m$\times70 \mu$m) and XRD measurements (blue bar, $20 \mu$m$\times 40 \mu$m), and measurement scan direction (gray arrow); Panel c) shows the spectroscopic reflectance at three specific locations on the lg-LSA stripe. Panel e) shows the reflectance spectroscopy heatmap (reflection coefficient mapped to color) with locations corresponding to those of panel c) indicated with vertical lines. Panel d) shows the XRD spectra (diffraction intensity vs. scattering vector, Q) at the same three locations on the lg-LSA stripe, and panel f) shows the corresponding XRD heatmap (diffraction intensity mapped to color).}
    \label{fig:mapcreation}
\end{figure*}

To demonstrate the value of the novel optical methods presented in Secs~\ref{sec:microscopy} and~\ref{sec:spectroscopy}, we apply them to study phase transitions of (meta)stable phases in two materials systems: (a) single-composition \ce{Ga2O3} and (b) composition-gradient La-Mn-O.

\ce{Ga2O3} has four known accessible crystalline polymorphs: $\alpha$, $\beta$, $\gamma$, and $\epsilon$~\cite{Roy1952,Yao2018,Stepanov2016} with distinct temperature regimes of thermodynamic stability, allowing clear interpretation of the phase space regions. $\beta$-\ce{Ga2O3} is of particular interest due to its commercial application in electronic and power devices as a wide bandgap semiconductor~\cite{Playford2013}. $\alpha$- and $\epsilon$-\ce{Ga2O3} polymorphs have, for example, applications in functional heterostructures, as transparent conductive oxides and exhibits tunable bandgaps~\cite{Yao2018,Stepanov2016}.

The La-Mn-O system has been the subject of experimental interest for many years due to its application in solid oxide fuel cells~\cite{Grundy2005} and the colossal magnetoresistance found in lanthanum manganite perovskites~\cite{vanRoosmalen1995,Jacob2003}.  The equilibrium pseudo-binary phase diagram in air was reviewed by Grundy~\textit{et al.}~\cite{Grundy2005} who identify four polymorphs of \ce{MnO_x}, two polymorphs of \ce{La2O3}, and the perovskite \ce{LaMnO3}. However, non-equilibrium metastable phases of the system have not been studied extensively. The richness of the chemical system along with interest in potential metastable phases makes it an excellent candidate for a combinatorial high-throughput study.

\subsection{Root mean square signal gradients\label{sec:RMSG}}
We commence our analysis of the data arising from the three different experimental sources by deriving key quantities that can be compared across the sampling techniques. The images and heat maps, as described in Sec.~\ref{sec:spatially}, share a common horizontal axis corresponding to the spatial location along the temperature-gradient (lateral) axis of a given lg-LSA stripe. Figure~\ref{fig:mapcreation} shows a visual comparison of the resulting images/maps for a representative sample from the La-Mn-O library, where we align the x-axis with respect to the center of the lg-LSA stripe. 

For a quantitative comparison we reduce the dimensionality of these images/maps by integrating along the y-dimension of the grayscale intensity in the microscopy images, the reflected photon energy in the spectroscopy maps, and the signal intensity in reciprocal space in the XRD maps. Note that for the microscopy images there is essentially no variation along the $y$ axis in the region imaged since the annealing conditions have reached steady state (i.e., along the longitudinal axis). Integration therefore simply averages over redundant information along that direction. 

The key quantity derived from these 1-D reduced datasets is the root mean square of the signal gradients (RMSG) with respect to the lateral position across a stripe. Regions with a high RMSG value correspond to rapid variations in the underlying measured properties associated with materials transformations. For the two optical methods, the physical source of these variations in the transparent films stems from varying interference due to changes in the film thickness $d$ and/or the refractive index $n$ of the film, and possible diffuse scattering of the light due to formation of a non-specular surface. In particular, small changes in either $n$ or $d$ markedly alter the optical reflectance of the material. Changes in the RMSG in the XRD patterns are directly associated to structural changes, including local ordering of the amorphous film and changes in crystalline phase of the material.

\begin{figure}[t]
    \centering
    \includegraphics[width=\columnwidth]{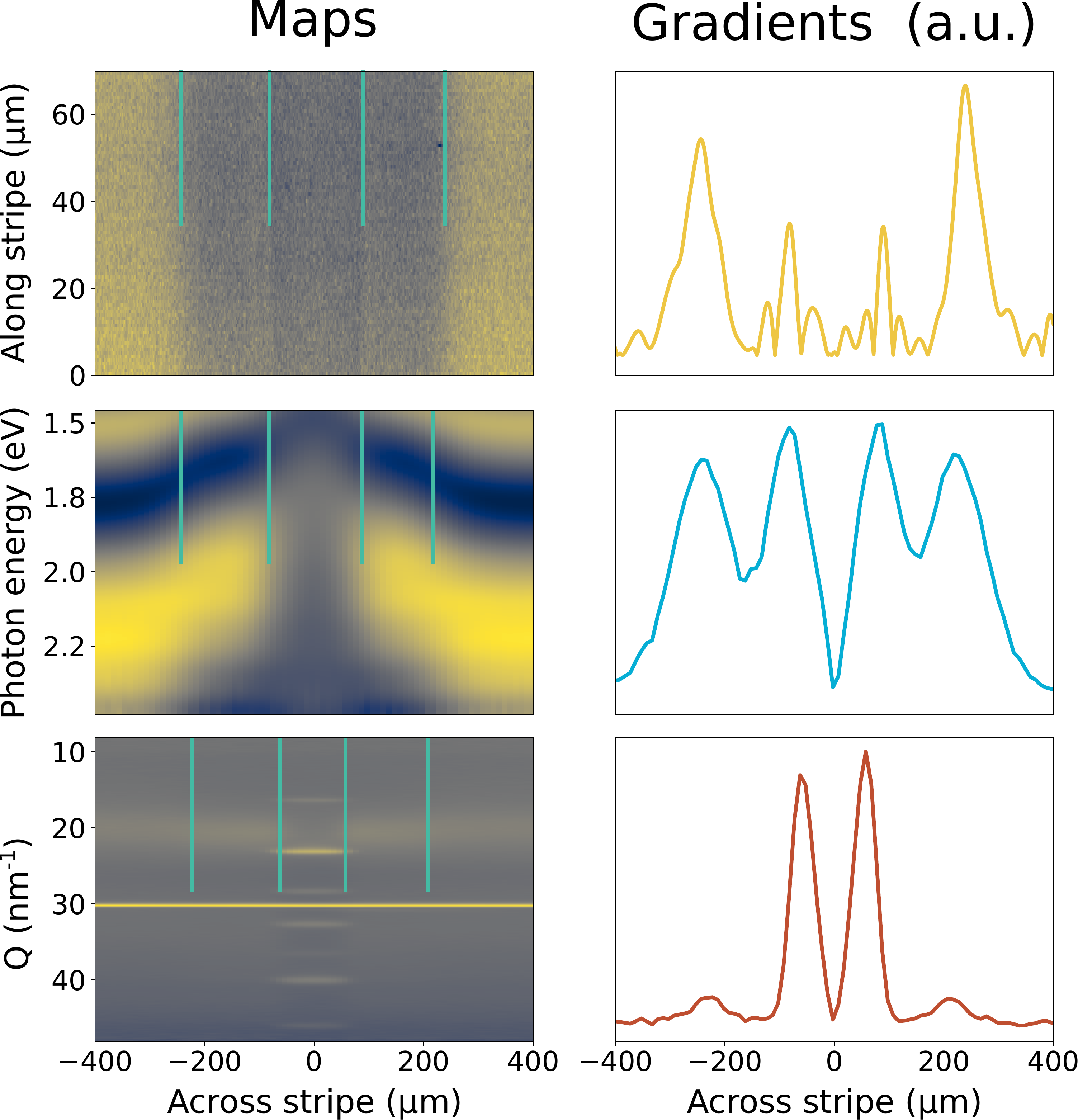}
    \caption{The optical micrograph, reflectance spectroscopy, and XRD maps aligned and stacked over the same spatial extent. The teal vertical lines indicating the peaks in the corresponding RMSG shown in the panels on the right. Note that the signal intensity along each column of the map is integrated for every spatial location before computing the lateral gradient.}
    \label{fig:compar}
\end{figure}

Figure~\ref{fig:compar} shows how the images/maps compare with each other for the same representative lg-LSA stripe as in Figure~\ref{fig:mapcreation}, and presents the corresponding RMSG in the right-hand panels. The height and widths of the peaks in the RMSG vary from technique to technique due to the sensitivity and  spatial resolution of each method. The location of the peaks in the RMSG are indicated by vertical teal lines in the signal maps; these align well across the three experimental methods. With the XRD patterns of known \ce{LaMnO3} phases ~\cite{Sayagues2012,Norby1995}, we can identify the  structural transformations in the film and associate them with the processing conditions on the stripe. As the maximum temperature experienced by the film ranges from room temperature at the position $x=-400 \ \mu$m towards the peak temperature of the entire lg-LSA stripe at the position $x=0 \  \mu$m, the film undergoes two definable transformations. First, the amorphous as-deposited \ce{La_{0.7}Mn_{0.3}O3} film begin to change to an optically denser condition $x=-250 \ \mu$m. Second, the fully dense amorphous film transforms to a crystalline cubic \ce{LaMnO3} structure at $x=-100 \ \mu$m. The symmetric nature of the lateral thermal gradient leads to the same set of transformations in reverse order on the right-hand half of the stripe.

Ideally, any materials transformation should be captured with all three methods, leading to a one-to-one mapping of the associated RMSG. In general the same symmetries and transformations are observed in the optical micrograph and the reflectance spectroscopy map, with RMSG peaks locations consistent with those observed in the XRD map. To quantify the difference of the RMSG across the techniques for all the experimental conditions, we use the Pearson correlation coefficient (PCC) between each pair of techniques as a measure of the similarity in the overall material transformation characterization. A PCC value of 0 equates to no linear correlation, whereas a value 1 or -1 equates to total positive or negative linear correlation, respectively. For the PCC calculation the RMSGs are interpolated with a cubic spline function spanning the same spatial extent. Note that the comparison is sensitive to the difference in spatial resolution of each technique (the effective probe sizes in Figure~\ref{fig:mapcreation}), which can cause a sharp transformation boundary to appear indistinct, or could create or obscure the appearance of a multi-phase region. Further, inaccurate calibration of the lateral anneal space can compress or expand the RMSG and significantly alter the calculated correlation coefficients. Hence, we carefully calibrated our spatial scales to minimize the effect of such artifacts.

\begin{figure*}[t]
\includegraphics[width=\textwidth]{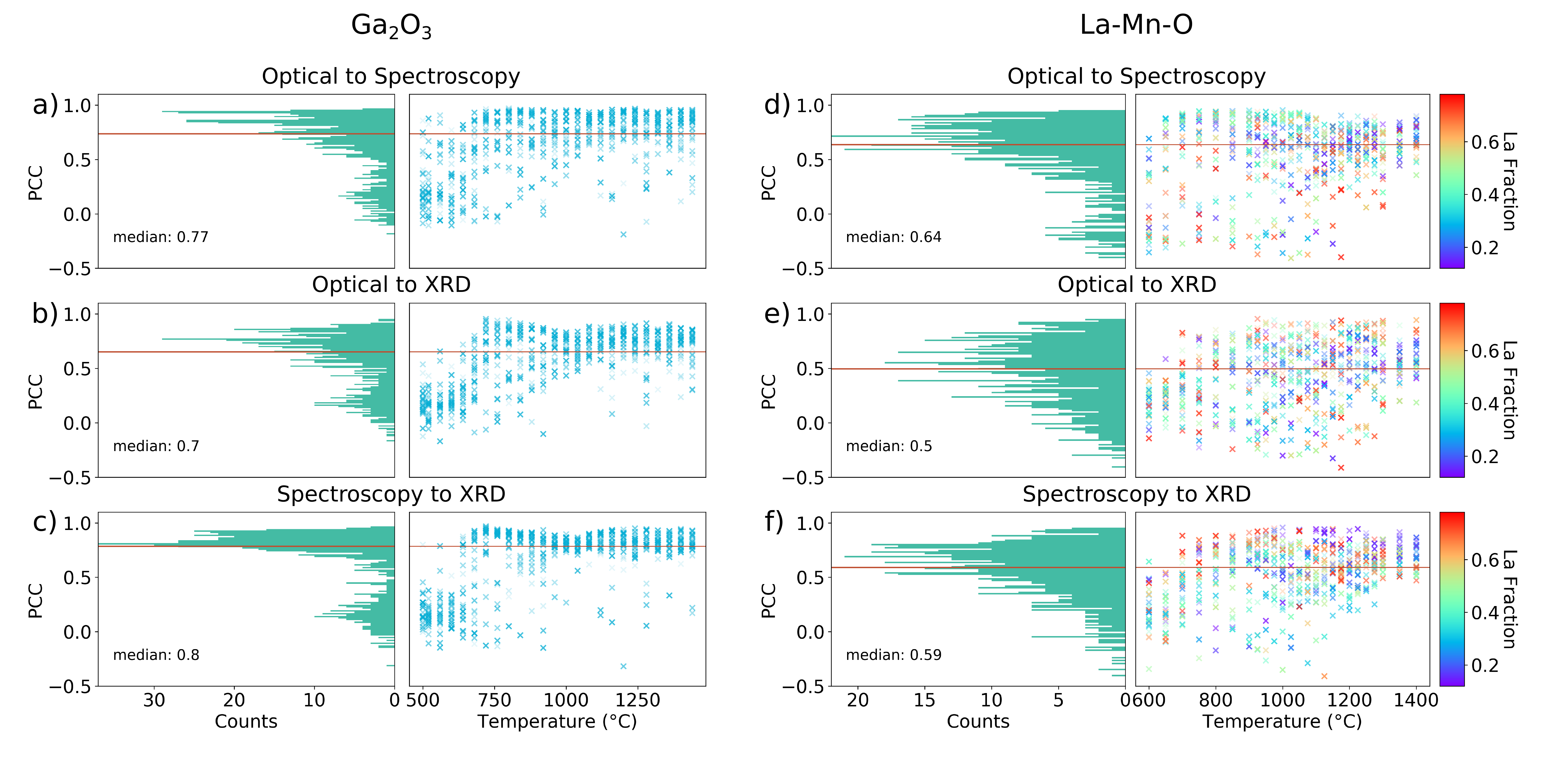}
    \caption{The PCCs comparing the RMSG between each technique across all peak temperature and dwell time conditions. Panels a), b), and c) show the histograms and temperature-sorted scatter plots of the PCCs for all LSA experiments conducted for the \ce{Ga2O3} sample, while panels d), e), and f) show the corresponding results for the La-Mn-O sample. The color gradient in the scatter plot corresponds to the La cation fraction.}
    \label{fig:LaMnOxStats}
\end{figure*}

The subplots a)--c) and d)--f) in Figure~\ref{fig:LaMnOxStats} summarize the PCC values between each pair of techniques (optical to spectroscopy, optical to XRD, spectroscopy to XRD) for the 612 and 603 LSA conditions analyzed in the \ce{Ga2O3} and La-Mn-O samples, respectively. The right-hand panels in each subplot show the scatter of PCC values between the specified techniques over all LSA conditions. For the \ce{Ga2O3} system, the variation in dwell conditions are encoded with different opacity (light and dark for short and long dwells, respectively), while for the La-Mn-O system the color encodes the compositional variation of the La/(La+Mn) atom fraction. The left panels in each subplot show the histogram of the PCC from the corresponding panel on the right. The red horizontal line denotes the median value of the PCC.

The histograms for the \ce{Ga2O3} system clearly show a bimodal distribution, which indicates the presence of distinct, systematic features in the PCCs over the investigated LSA conditions. The first mode exhibits a peak with a median of $\approx 0.1$, while the peak of the second mode is located at a value of $\approx 0.7-0.8$, with some variation depending on the compared techniques. A comparison with the corresponding scatter plot shows that the two modes stem from samples that are separated distinctly at a critical temperature $T_{crit} \approx 600^{\circ}$C that is independent of the anneal time as observed in Figure~\ref{fig:LaMnOxStats}~a)--c). We ascribe this sharp increase in the PCC to the onset of crystallization from the amorphous \ce{Ga2O3} film, consistent with the findings reported previously~\cite{Zhang2014}. Below the threshold value $T_{crit}$, there is essentially no variation in the RMSG since the signal stems from purely amorphous samples, leading to correlation values close to zero. Above $T_{crit}$, the crystallization events lead to clear changes in the RMSG that correlate strongly across the techniques. Note that the correlations between the spectroscopy and XRD characterization has the tightest distribution for the second mode (above $T_{crit}$), which we attribute to the very similar spatial resolutions of the two techniques.

A similarly sharp onset is neither expected nor observed for the La-Mn-O system due to the increased complexity associated with compositional variation across the wafer. The PCC gradually increases as a function of anneal temperature between 500$^{\circ}$C and 800$^{\circ}$C, leading to a unimodal, skewed distribution in the histograms. The median value is $\approx 0.6$, which is lower than the second mode of the \ce{Ga2O3} system but still indicates a strong correlation. Thin films of \ce{La2O3} and \ce{MnO2} crystallize from amorphous as-deposited films at 600$^{\circ}$C and 450$^{\circ}$C \cite{Fau1994,Jun2003}, respectively. While various lanthanum manganate materials have been investigated, they typically incorporate various substitutions \cite{Nokhrina1990} that complicate comparative measurements, or are grown on a heated substrate at 900$^{\circ}$C to synthesize as-deposited crystalline films \cite{Kim2010}.

Overall, we observe a high degree of correlation ($>0.5$) in the PCC analysis, which clearly indicates that the optical and spectroscopic measurements are sensitive enough to identify the material transformations observed by XRD. For both materials systems, the strongest correlation is observed when comparing spectroscopy to XRD. This result is not surprising, since these two methods contain a broader spectrum of information (photon energy and diffraction angles) compared to microscopy (single value from the three color channels). Nevertheless, microscopy imaging is a simple and powerful method to rapidly assess the RMSG features since the timescale to capture a single RGB image is 2-3 orders of magnitude shorter than that needed for reflectance spectroscopy or XRD (when available). 

\subsection{Completeness of the material transformations\label{sec:completeness}}

\begin{figure}[b]
\includegraphics[width=\columnwidth]{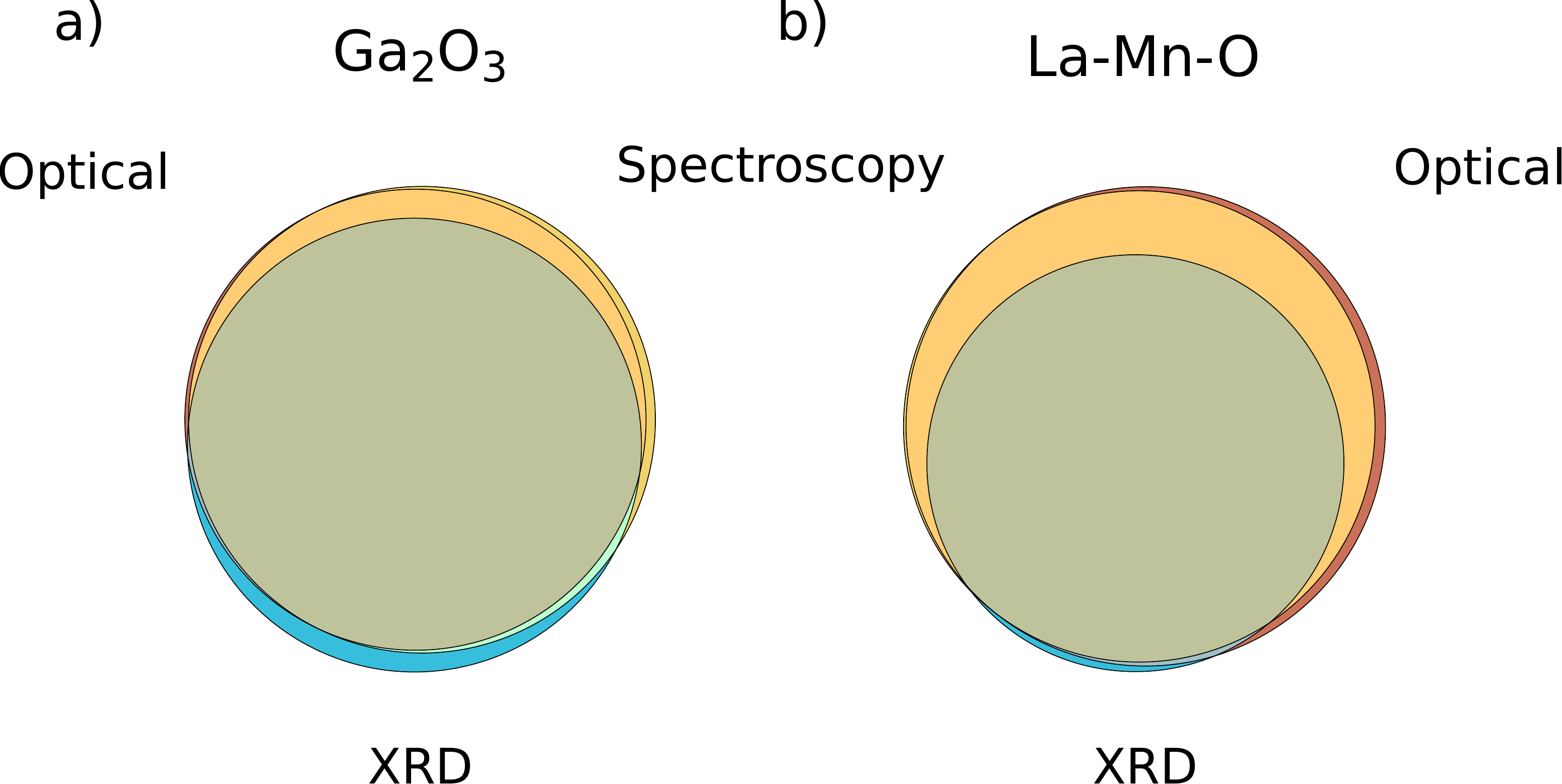}
    \caption{Venn diagrams illustrating the intersections between all three techniques for the  La-Mn-O and  \ce{Ga2O3} samples in the panels a) and b), respectively. The optical microscopy, reflectance spectroscopy and XRD areas are colored in red,  yellow and blue, respectively. False negatives are represented by as small blue slivers at the bottom of each Venn diagram.}
    \label{fig:VennFig}
\end{figure}

The PCC serves as a statistical method to assess and compare the RMSG, but does not reflect whether or not the optical methods are sensitive to a particular type of crystallographic transformation occurring in the LSA processed thin films, i.e., it does not clarify what \textit{kind} of transformations are identified or missed. To address this issue, we employ a metric complementary to the PCC by analyzing the completeness of the material transformations identified by XRD to those identified by either of the  optical techniques. Of particular concern are the cases where crystalline structure changes occur (based on the complete XRD data) without detectable changes in the optical techniques, i.e., ``false-negatives''. To assess the presence of such cases, we manually compiled a list of transformations detected in any or all sets of techniques in a given stripe through visual inspection and identified the false negatives.
``False-positives'' correspond to transitions observed in either of the optical methods corresponding to putatuve phase transformations that do not appear in the XRD measurements. The completeness of observed structural transformations are best visualized with a Venn diagram (Figure~\ref{fig:VennFig}). The subsets represent the portion of the transformations identified by one, two, or all three characterization techniques. The total number of identified transformations, independent of transformation type, within each set are listed in Table~\ref{tbl:VennSubsets}.

In contrast to what one would expect from the PCC analysis, the anneal stripes in the \ce{Ga2O3} system display very gradual materials transformations across the stripe that lead to a relatively large false negative rate of approximately 4.3\% for each of the optical techniques. This value is substantially higher than for the La-Mn-O system, but is low on an absolute scale. A detailed investigation of this subset of false negatives reveals that they occur primarily when the $\gamma$ phase transforms to the $\beta$ phase with increasing temperature. This result confirms that the two optical methodologies may be insensitive to certain types of materials transformation, particularly when it occurs gradually between 2 phases with similar optical properties. Both optical microscopy and reflectance spectroscopy identified almost identical sets of material transformations, i.e., the number of transformations \textit{only} observed in microscopy ($\text{O}\setminus (\text{S}\cup \text{X})$)) and \textit{only} observed in spectroscopy ($\text{S}\setminus (\text{O}\cup \text{X})$) are very small. The total  false positive rate for the \ce{Ga2O3} sample, i.e., $(\text{S}\cup \text{O})\setminus \text{X}$, is 9.8$\%$.

\begin{table*}
\small
  \caption{Venn diagram values for material transformations identified by the optical microscopy (O), reflectance spectroscopy (S), and/or XRD (X) techniques. The critical false negatives are $\text{X}\setminus (\text{O}\cup \text{S})$, while false positives leading to unnecessary XRD measurements are $(\text{S}\cup \text{O})\setminus \text{X}$.}
  \label{tbl:VennSubsets}
  \begin{ruledtabular}
\begin{tabular}{|c|c|c|c|c|c|c|c|}

Library & $\text{S}\cup\text{O}\cup \text{X}$ & $(\text{S}\cup \text{O})\setminus \text{X}$& $(\text{O}\cup \text{X})\setminus \text{S}$&$(\text{S}\cup \text{X})\setminus\text{O}$ &
$\text{O} \setminus(\text{S}\cup \text{X})$ & $\text{S}\setminus (\text{O}\cup \text{X})$ & $\text{X}\setminus (\text{O}\cup \text{S})$\\

 \hline
\ce{Ga2O3} & 848 & 78 & 5 & 12 & 2 & 18 & 39 \\
\hline
La-Mn-O & 943 & 279 & 8 & 2 & 39 & 3 & 7\\

\end{tabular}
\end{ruledtabular}
\end{table*}

For the La-Mn-O system, 0.7$\%$ of the structural transformations are demonstrably false negatives ($\text{X}\setminus (\text{S}\cup \text{O})$; see tiny blue sliver at bottom of Figure~\ref{fig:VennFig} b)). The sensitivities of each optical method applied to the La-Mn-O system are similar to those for \ce{Ga2O3} with regard to capturing identical material transformations. The La-Mn-O system shows a larger false positive rate, 25.1\%,  higher than that in \ce{Ga2O3}. In general, such false positives can be attributed to dust, delamination, or other defects in the thin film that do not pertain to the underlying crystallographic transformations.

As method for pre-screening phase transitions prior to XRD experiments, both optical methods are effective despite the large number of false positives. Such false positives are undesirable because they motivate redundant sampling but they do not lead to any loss of information about crystallographic material transformations. For example, exhaustive XRD sampling would require ~125,000 individual measurements in the La-Mn-O composition spread, while focussing only spectroscopy-detected transitions would reduce the the number of required measurements to as few as 960. Thus the false positive rate of 23\% nevertheless enables substantial resource savings in terms of the total number of required XRD measurements.

On the other hand, false negatives may result in loss of information in the synthesis phase diagram and should be minimized. The low rate false-negative rate of approximately 0.7\% in the La-Mn-O sample demonstrates the capability to effectively pre-screen transition boundaries prior to XRD experiments. The higher false-negative rate of 4.3\% observed in the \ce{Ga2O3} sample stems from a single type of phase transformation that  is especially difficult to detect optically, i.e., in which the optical absorption and optical thickness (the product of $n$ and $d$) both remain approximately constant. Such instances are coincidental and expected to be rare in higher order composition spaces.

\subsection{Accelerating experimental throughput}

To assess the value of this work in accelerated exploration of materials synthesis spaces, we consider the demonstrated workflow of i) composition library deposition, ii) LSA-based processing with approximately 600 LSA stripes (peak temperature and dwell time conditions) per composition library, and iii) characterization within each LSA stripe to identify transitions in temperature, dwell time, and composition.

The throughput of these processes in terms of number of composition libraries is 1 and 10 per day for deposition and lg-LSA, respectively. For transition characterization, the alternate approaches described in the present work -- optical microscopy, reflectance spectroscopy, and $\mu$m-resolution XRD -- have throughputs of 1000, 2, and 10 per day, respectively, although the XRD throughput can be substantially lower for thin films that require larger acquisition times due to weak scattering. 

With these nominal throughputs, the deposition step appears to be approximately rate-limiting until one considers resource availability, which is of order of 9 days per year for the specialized synchrotron XRD measurement and 300 days per year for the other purpose-built optical instruments. Limited XRD availability makes this technique rate-limiting for annual workflow throughput, and this limitation is exacerbated when considering the value of feedback--the ability to choose the deposition conditions based on consideration of all prior data.

Typical synchroton XRD throughput is up to 90 libraries per year corresponding to 90 traversals of the workflow, although deposition selection informed by characterization of all prior libraries only occurs in up to 9 of those traversals because during the 9 days of XRD experiments; decision throughput is limited by deposition throughput. The ability to detect transitions with optical techniques increases the decision throughput from 9 to 300 per year, enabling optimized allocation of the XRD resource to structural characterization of the optically-identified transitions.

This scheme is particularly effective if the XRD-based (structural) transitions are consistently identified with the optical techniques; Figure $\ref{fig:VennFig}$ provides the most extensive demonstration to date that optically-detected transitions are most often due to structural transitions. While the throughput of the optical microscopy technique is much higher than that of reflectance spectroscopy, in the context of the annual workflow analysis the increased throughput is not the essential benefit. The value of the microscopy lies in its spatial resolution. The spectral resolution of reflectance spectroscopy facilitates transition detection but its spatial resolution is limited in practice to $\approx 5 \mu$m. If multiple transitions occur within that length scale (which corresponds to a range in temperature of up to $\approx 25$~K for our lg-LSA system), optical microscopy becomes a critical complementary transition detector. Narrowly clustered transitions were not observed in the data presented herein, but the concurrence of the transitions identified by microscopy and spectroscopy indicate that the combination of these techniques provides a robust approach to transition detection in synthesis parameter space. 

\section{Conclusions\label{sec:conclusions}}

The methods described here enable high-throughput experimental identification of materials phase transitions across an lg-LSA stripe. Using rapid, low-cost, optical techniques, such as optical microscopy and spatially resolved reflectance spectroscopy, locations of materials transitions can be identified and can serve as prior knowledge for experimentally expensive techniques such as synchrotron XRD studies.  For both single-composition and composition-gradient samples, structural transitions were identified via optical methods. An anomalous transition behavior was noted in \ce{Ga2O3}, where a structural change -- the introduction and growth of a secondary phase -- was not always identified via optical means, demonstrating the possibility of false negatives under certain circumstances.  The appearance of unidentified transitions and the common occurrence of optical changes not corresponding to a structural change (false positives) suggest that further development could improve the effectiveness of the methods and extends the properties that can be identified optically.  Collectively, the results demonstrate the utility of using in-lab optical characterization methods to accelerate iterative feedback between processing and thin film deposition in the discovery and development of new materials. The results also motivate further development of algorithms that leverage the optical characterization techniques to design synchrotron XRD experiments that enhance the efficient utilization of that limited resource.

\section{Acknowledgements}

The authors acknowledge the Air Force Office of Scientific Research for support under award FA9550-18-1-0136. This work is based upon research conducted at the Materials Solutions Network at CHESS (MSN-C) which is supported by the Air Force Research Laboratory under award FA8650-19-2-5220. This work was also performed in part at the Cornell NanoScale Facility, a member of the National Nanotechnology Coordinated Infrastructure (NNCI), which is supported by the National Science Foundation (Grant NNCI-1542081). MA acknowledges support from the Swiss National Science Foundation (project P4P4P2-180669).



%
\end{document}